\documentclass[usenatbib,longtabl]{mn2e}
\input psfig.sty

\title[The dynamical stability of W UMa systems]
{The dynamical stability of W Ursae Majoris-type systems}

\author[L. Li and  F. Zhang]
{Lifang Li \thanks{E-mail:
gssephd@public.km.yn.cn or llf@ynao.ac.cn}, and Fenghui Zhang\\
National Astronomical Observatories/Yunnan Observatory, Chinese Academy of
Sciences, P.O.Box 110,\\
Kunming, Yunnan Province 650011, P. R. China
}

\begin{document}

\date{Accepted yy mm dd. Received yy, mm, dd; in original form 2006 February 13}

\pagerange{\pageref{firstpage}--\pageref{lastpage}} \pubyear{2006}

\maketitle

\label{firstpage}

\begin{abstract}
Theoretical study indicates that a contact binary system would
merge into a rapidly rotating single star due to tidal instability
when the spin angular momentum of the system is more than a third
of its orbital angular momentum. Assuming that W UMa contact
binary systems rigorously comply with the Roche geometry and the
dynamical stability limit is at a contact degree of about 70\%, we
obtain that W UMa systems might suffer Darwin's instability when
their mass ratios are in a region of about 0.076--0.078 and merge
into the fast-rotating stars. This suggests that the W UMa systems
with mass ratio $q\leq0.076$ can not be observed. Meanwhile, we
find that the observed W UMa systems with a mass ratio of about
0.077, corresponding to a contact degree of about 86\% would
suffer tidal instability and merge into the single fast-rotating
stars. This suggests that the dynamical stability limit for the
observed W UMa systems is higher than the theoretical value,
implying that the observed systems have probably suffered the loss
of angular momentum due to gravitational wave radiation (GR) or
magnetic stellar wind (MSW).

\end{abstract}


\section{Introduction}

Interest in W UMa binaries was revived recently with the discovery
of large number of new ones among blue stragglers in open and
globular clusters \citep{kal88,kal90,mat90,yan94,mat96}. It
appears likely that at least some of blue stragglers are formed by
the merging of contact binaries. The models of W UMa systems
\citep{li04,li05} had shown that W UMa systems evolve into contact
binaries with extreme mass ratios and then evolve into single,
fast-rotating stars (FK Com stars) or blue stragglers due to
Darwin's instability when the spin angular momentum of the system
is more than a third of its orbital angular momentum
\citep{hut80,egg01}.

The spin angular momentum has a significant influence on the
evolution of W UMa contact binaries, especially for those with
extreme mass ratios. Neglecting the spin angular momentum of the
small secondary, \citet{ras95} first derived a cutoff mass ratio
for contact binary systems and the cutoff mass ratio depends on
the structure and the contact degree. Meanwhile, \citet{ras951}
identified the dynamical stability limit at a contact degree of
about 70\% for W UMa systems. \citet{yan051} derived a cutoff mass
ratio of about 0.021 from the observed W UMa systems with mass
ratios $q<0.25$. In this work, using the observational data and
the theoretical results reported recently, we investigate the
dynamical stability limit of W UMa systems, and find that W UMa
systems begin merging when their mass ratios are in a region of
about 0.076--0.078. This suggests that W UMa systems with mass
ratio $q\leq0.076$ can not be observed. Meanwhile, we estimate
that the observed W UMa systems with a minimum mass ratio of about
0.077, corresponding to a contact degree of about 86\%, would
suffer Darwin's instability and begin merging. This suggests that
the minimum mass ratio for the observed W UMa systems is larger
than the theoretical value, implying that the observed W UMa
systems have suffered the loss of angular momentum loss owing to
GR or MSW.

\section{Dynamical Stability Limit}

It is well known that a contact binary would suffer Darwin's
instability and then coalesce into a fast-rotating single star
(including FK Com-type stars and blue stragglers) when the spin
angular momentum of the system is more than a third of its orbital
angular momentum. The spin angular momentum of a binary star can
be expressed as
\begin{equation}
J_{\rm spin} = (k_{1}^{2}M_{1}R_{1}^{2} +
k_{2}^{2}M_{2}R_{2}^{2})\omega_{\rm s},
\end{equation}
where $M_{1,2}$ and $R_{1,2}$ are the masses and radii of the
primary and the secondary in solar units, $k_{1,2}$ the ratios of
the gyration radii to the stellar radii for both components, and
$\omega_{\rm s}$ the spin angular velocity. The orbital angular
momentum of the binary system reads

\begin{equation}
J_{\rm orb} = \frac{M_{1}M_{2}}{M_{1}+M_{2}}A^{2}\omega_{\rm o},
\end{equation}
where $\omega_{\rm o}$ is the orbital angular velocity and $A$ the
orbital radius of the binary. We assume that the W UMa systems are
in synchronous rotation (i.e. $\omega_{\rm s}=\omega_{\rm o}$) and
the ratios of gyration radii to the stellar radii for both
components are equal (i.e. $k_{1}^{2}=k_{2}^{2}=k^{2}$), the ratio
of the spin angular momentum to the orbital angular momentum can
be written as

\begin{eqnarray}
R&=&\frac{J_{\rm spin}}{J_{\rm orb}} =\frac{k^{2}(M_{1}R_{1}^{2} +
M_{2}R_{2}^{2})}{A^{2}M_{1}M_{2}/(M_{1}+M_{2})} \cr
&=&k^{2}\frac{1+q}{q}\Bigl(\frac{R_{1}}{A}\Bigr)^{2}\Biggl(1+q\Bigl(\frac{R_{2}}{R_{1}}\Bigr)^{2}\Biggr),
\end{eqnarray}
where $q$ ($=M_{2}/M_{1}<1$) is the mass ratio of the binary
system. If a contact binary system is a marginal contact one in
which two components has just filled the inner Roche lobes,
following \citet{egg83}, the relative radii of both components of
the contact binary read
\begin{equation}
\frac{R_{2}}{A}=\frac{0.49q^{2/3}}{0.6q^{2/3}+{\rm
ln}(1+q^{1/3})},
\end{equation}
\begin{equation}
\frac{R_{1}}{A}=\frac{0.49q^{-2/3}}{0.6q^{-2/3}+{\rm
ln}(1+q^{-1/3})}.
\end{equation}
If the two components of a contact binary system have filled the
outer Roche lobes, the relative radii of the components can be
expressed by the following equations \citep{yak05}
\begin{equation}
\frac{R_{2}}{A}=\frac{0.49q^{2/3}+0.27q-0.12q^{4/3}}{0.6q^{2/3}+{\rm
ln}(1+q^{1/3}) },
\end{equation}
\begin{equation}
\frac{R_{1}}{A}=\frac{0.49q^{-2/3}+0.15}{0.6q^{-2/3}+{\rm
ln}(1+q^{-1/3}) }.
\end{equation}
Using Eqs. (3), (4) and (5) and taking $k^{2}$ to be the values of
main sequence stars (from 0.05 for no convection stars to 0.21 for
fully convective stars), we can obtain a relationship between the
angular momentum ratio $R_{\rm i}$ and the mass ratio $q$ for
marginal contact systems
\begin{eqnarray}
R_{\rm i}
=k^{2}\frac{1+q}{q}\Bigl(\frac{0.49q^{-2/3}}{0.6q^{-2/3}+{\rm
ln}(1+q^{-1/3})}\Bigr)^{2} \times \cr
\Biggl(1+q\Bigl(\frac{0.6q^{2/3}+q^{4/3}{\rm
ln}(1+q^{-1/3})}{0.6q^{2/3}+{\rm ln}(1+q^{1/3})}\Bigr)^{2}\Biggr),
\end{eqnarray}
setting equation $R_{\rm i}=1/3$, and solving (numerically) for
$q$, we obtain the minimum mass ratio $q_{\rm min,in}$ for
stability of marginal contact binaries with the different values
of $k^{2}$ of MS stars and the relation between the minimum mass
ratio $q_{\rm min,in}$ and $k^{2}$ is shown in Figure 1.  Using
Eqs. (3), (6) and (7), we obtain a relationship between the
angular momentum ratio $R_{\rm o}$ and the mass ratio $q$ for
contact systems in which both components have filled the outer
Roche lobes
\begin{eqnarray}
R_{\rm o}
=k^{2}\frac{1+q}{q}\Bigl(\frac{0.49+0.15q^{\frac{2}{3}}}{0.6+q^{\frac{2}{3}}{\rm
ln}(1+q^{-1/3})}\Bigr)^{2} \Biggl(1+ q\times \cr
\Bigl(\frac{(0.49+0.27q^{\frac{1}{3}}-0.12q^{\frac{2}{3}})[0.6+q^{\frac{2}{3}}{\rm
ln}(1+q^{-\frac{1}{3}})]}{(0.49q^{-\frac{2}{3}}+0.15)[0.6q^{\frac{2}{3}}+{\rm
ln}(1+q^{\frac{1}{3}})]}\Bigr)^{2}\Biggr),
\end{eqnarray}
setting equation $R_{\rm o}=1/3$, and solving (numerically) for
$q$, we obtain the minimum mass ratio, $q_{\rm min,out}$, for
stability of W UMa systems in which two components have filled the
outer Roche lobes with the different values of $k^{2}$ and the
relation between $q_{\rm min,out}$ and $k^{2}$ is also shown in
Figure 1. As seen from Figure 1, the value of $q_{\rm min,in}$ is
smaller than that of $q_{\rm min,out}$ at any value of $k^{2}$.
This suggests that the stability of a contact binary with given
masses indeed depends on the degree of contact $f$, or filling
factor, $0<F=1-f<1$. Meanwhile, the more is the value of $k^{2}$,
the larger are the minimum mass ratios ($q_{\rm min,in}$ and
$q_{\rm min,out}$) and the difference between $q_{\rm min,in}$ and
$q_{\rm min,out}$.

\begin{figure}
\centerline{\psfig{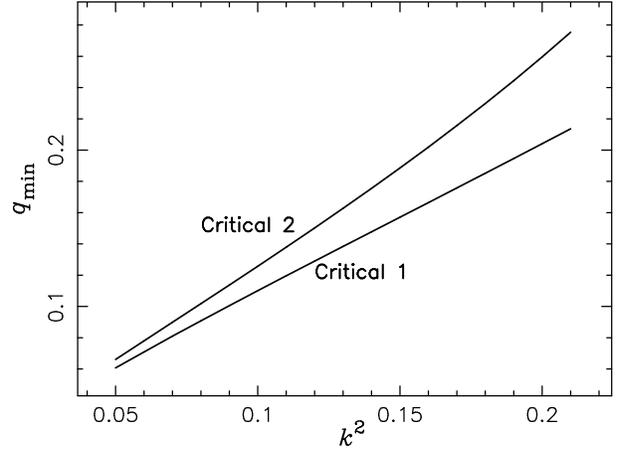}}
\caption{The minimum mass ratio for stability $q_{\rm min}$ in the
Roche approximation. The parameter $k^{2}$ is the dimensionless
gyration radius of the stars} \label{fig1}
\end{figure}

\begin{figure}
\centerline{\psfig{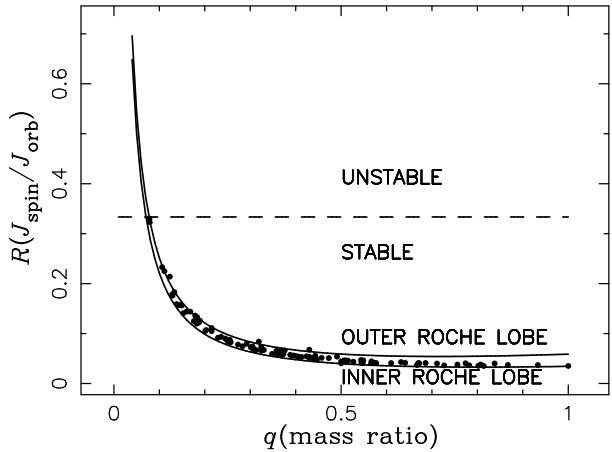}}
\caption{The angular momentum ratio $R$ against the mass ratio,
$q$. The solid lines represent theoretical results corresponding
to the systems filling the inner Roche lobe and the outer Roche
lobe, respectively. The solid dots represent the observed W UMa
systems. } \label{fig2}
\end{figure}

\begin{figure}
\centerline{\psfig{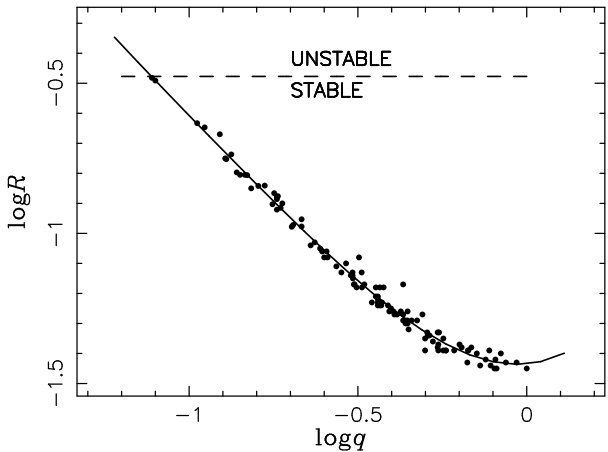}}
\caption{The angular momentum ratio $R$ against the mass ratio,
$q$. The solid line represents the fitting result. The solid dots
represent the observed W UMa systems. } \label{fig3}
\end{figure}

We take $k^{2}=0.06$ as \citet{ras95}, the ratio of the spin
angular momentum to the orbital angular momentum against the mass
ratio is plotted in Figure 2. Meanwhile, we obtain $q_{\rm
min,in}=0.071$ and $q_{\rm min,out}=0.078$ for $k^{2}=0.06$. In
general, the dynamical instability of W UMa contact binary systems
should be occurred at an overcontact stage with a higher degree of
contact rather than at a marginal contact one. The numerical study
of the equilibrium and stability of close binary systems
\citep{ras951} had identified the dynamical stability limit at a
contact degree of about 70\%, corresponding to a minimum mass
ratio of about 0.076. This suggests that the W UMa systems with
mass ratios in a region of 0.076--0.078 would suffer Darwin's
instability and then coalesce into the fast-rotating single stars
(including FK Com stars and blue stragglers), implying that the W
UMa systems with mass ratios $q\leq0.076$ can not be observed. We
collect the absolute parameters of some of W UMa systems (listed
in Table 1) reported recently, together with those compiled by
\citet{mac96}, \citet{gaz05}, and by \citet{yak05}, to determine
the minimum mass ratio of the observed W UMa systems. Taking
$k^{2}=0.06$ as mentioned above, the observed systems are also
plotted in Figure 2. It is seen in Figure that, except for two
systems with a larger scatter, all systems load in the region
limited by the theoretical curves corresponding to the inner and
outer Roche lobes, and none of them has surpassed the dynamical
stability limit. However, it is difficult to obtain a direct
polynomial relation between the ratio $R$ and the mass ratio $q$
by the least-squares fitting. But we have found a relation between
the logarithms of the ratio $R$ and the mass ratio $q$ as the
following
\begin{eqnarray}
{\rm log}R=-1.4347+0.09954{\rm log}q +1.8506({\rm log}q)^{2} \cr
+1.25287({\rm log}q)^{3}+0.3291({\rm log}q)^{4},
\end{eqnarray}
the observed systems, together with the theoretical curve, are
shown in Figure 3. As seen from Figure 3, the theoretical curve
has given a good fit to the observed systems. Setting equation
${\rm log}R={\rm log}(1/3)$, and solving (numerically) for $q$, we
obtain a minimum mass ratio, $q_{\rm min,obs}=0.077$, for the
observed systems. It is larger than 0.076 estimated from the
theoretical result \citep{ras951}. This suggests that the observed
W UMa systems have probably suffered the loss of angular momentum.
Since the loss of angular momentum of the observed systems would
result in the shrinking of their orbits and thus would cause their
inner and outer Roche lobes to decrease and contact degree $f$ to
increase, so that leads $f$ to have exceeded 70\% before the
observed systems reaching the minimum mass ratio $q=0.076$.

\begin{table}
\begin{footnotesize}
Table~1.\hspace{4pt} The physical parameters of some W UMa systems.\\
\begin{minipage}{17cm}
\begin{tabular}{l|cccccc}
\hline\hline\\
{Stars}&{$P$}&{$M_{1}$}&{$q$}&{$R_{1}$}& {$R_{2}$}&{Refs.}\\

&{(days)}&{($M_{\rm \odot}$)}& &{($R_{\rm
\odot}$)}&{($R_{\rm \odot}$)} &\\
\hline

V410 Aur  & 0.3663&1.30 & 0.146 &   1.40& 0.61& (1) \\
XY Boo & 0.3705& 0.91& 0.186& 1.23& 0.61&(1)\\
AH Cnc & 0.3605 &1.21 & 0.149 &  1.36 & 0.62&(2) \\
CW Cas & 0.3188 &1.06 & 0.547 & 1.01 &0.76& (3)\\
V776 Cas &0.4404 &1.63 & 0.13& 1.71 & 0.71& (4) \\
V899 Her & 0.4212&2.10 &0.566 & 1.57 &1.22& (5) \\
BB Peg & 0.3615  &1.38 &0.362 & 1.26 & 0.76&(6) \\
V351 Peg& 0.5933&1.63&0.361&1.87&1.19&(7)\\
AU Ser & 0.3865& 0.90& 0.710& 1.10& 0.94& (8)\\
VZ Psc& 0.2613& 0.81 & 0.800& 0.78 & 0.70& (9)\\
\hline
\end{tabular}
\end{minipage}
\end{footnotesize}\\
{References in Table 1:(1) Yang et al.(2005a); (2) Zhang et
al.(2005); (3)Barone et al.(1988); (4) Djurasevic(2004);
(5)Ozdemir et al. (2002); (6) Lu \& Rucinski(1999); (7) Albayrak
et al.(2005);  (8) Gurol(2005); (9) Hrivnak et al.(1995)}
\end{table}

\section{Discussions and conclusions}

The models of low-mass W UMa-type systems \citep{li04,li05} have
shown that W UMa systems would suffer Darwin's instability and
evolve into fast-rotating single stars (i.e. FK Com stars or blue
stragglers). \citet{ras95} predicted a cutoff of the mass ratio
for W UMa systems at about 0.09. Considering the rotation of the
secondary, we obtain that the minimum mass ratio for stability of
the marginal contact binaries is of about $q_{\rm min in}=0.071$,
suggesting that the rotation of the secondary still has a
definitely influence on the determination of the dynamical
stability limit of W UMa systems. \citet{ras95} had predicted that
the dynamical stability limit should depend on the fill-factor of
W UMa systems. Using the new results in studies of the W UMa
systems, we investigate the dynamical stability limit of W UMa
systems in which both components fill their outer Roche lobes and
obtain a minimum mass ratio $q_{\rm min,out}$ of about 0.078 for
this kind of systems. This suggests that the dynamical stability
limit of W UMa systems indeed depends on their fill-factor.

The difference between $q_{\rm min,in}$ and $q_{\rm min,out}$
suggests that the dynamical stability of the W UMa contact
binaries with given mass depends on the degree of contact
[$f=(\Omega-\Omega_{\rm in})/(\Omega_{\rm out}-\Omega_{\rm in}$)],
which is consistent with the result predicted by \citet{ras95}.
\citet{ras951} had identified the dynamical stability limit of W
UMa systems at about $f=70\%$, which corresponds to a minimum mass
ratio of about 0.076. We obtain a minimum mass ratio of about
0.077 , corresponding to a contact degree $f=(q_{\rm
min,obs}-q_{\rm min,in})/(q_{\rm min,out}-q_{\rm
min,in})\approx86\%$, for the observed systems. The difference
between the Rasio \& Shaprio's result and ours derived from the
observed systems suggests that the observed W UMa systems have
probably suffered the loss of angular momentum owing to GR or MSW
during the evolution of the observed W UMa systems. In fact, {\it
Einstein} soft X-ray observations and {\it International
Ultraviolet Explorer (IUE)} Ultraviolet observations \citep{vai80,
eat83,ruc83,vil83} have shown that W UMa systems are strong
sources. This suggests surface activity of the kind observed on
our Sun, and so presence of the magnetic field. Since the observed
W UMa systems rapidly lose the angular momentum via MSW, the
contact degrees of them have exceeded 70\% before their mass
ratios decrease to the minimum mass ratio q=0.076.

Among the about 200 contact binaries with reliable photometric
data, the system V857 Her has an extremely small mass ratio,
$q=0.0653$ \citep{qia05} for observed W UMa systems at present.
This is not only a new "record" for contact binary, surpassing the
well-known AW UMa with $q=0.075$ \citep{ruc92} and SV Crv with
$q=0.066$ \citep{ruc01}. It is a tremendous contribution for
theory that W UMa systems with the extremely small mass ratios are
found. Only a slight difference between observations and theory
not only indicates that theory have come close to a decent
prediction, but also indicates that some physical processes taking
place in the observed W UMa systems have not be considered in the
present theory. At first, the dynamical stability limit of W UMa
systems may also depend on the structure (i.e. the value of
$k^{2}$) and the value of $k^{2}$ may decrease with the evolution.
In addition, the differential rotation of the observed systems
would lead the ratio of the gyration radii to the stellar radii of
the components of the W UMa systems to be decreased. Therefore,
the differential rotation of the components of the W UMa systems
is another physical reason why some W UMa systems with the mass
ratios smaller than the minimum mass ratio 0.076 still can be
observed. However, contact binaries with mass ratios significantly
below that of SX Crv and V857 Her should not be observed, since
they are tidally unstable and quickly merge into a single,
fast-rotating object, on a tidal timescale of about
$10^{3}$--$10^{4}$ yr \citep{ras95}.

\section*{ACKNOWLEDGEMENTS}

This work was partly supported by the Chinese Natural Science
Foundations (Nos. 10273020, 10303006, and 10433030), the Yunnan
Natural Science Foundation (Grant No. 2005A0035Q), and by the
Foundation of Chinese Academy Sciences (KJCX2-SW-T06). We would
like to thank the referee Prof. F. A. Rasio for his helpful
comments and suggestions which greatly improved this paper.

\bsp

\label{lastpage}


\begin{thebibliography}{}
\bibitem[\protect\citeauthoryear{Albayrak et al.}{2005}]{alb05} Albayrak B., Djurasevic G.,
Selam S. O., et al., 2005 New Astron., 10, 163
\bibitem[\protect\citeauthoryear{Barone et al.}{1988}]{bar88} Barone F., Milano L.,
Maceroni C., \& Russo G., 1988, A\&A, 197, 347
\bibitem[\protect\citeauthoryear{Djurasevic et al.}{2004}]{dju04} Djurasevic G., Albayrak
B., Selam S. O., et al., 2004, New Astron., 9, 425
\bibitem[\protect\citeauthoryear{Eaton}{1983}]{eat83} Eaton J. A.,
1983, ApJ, 268, 800
\bibitem[\protect\citeauthoryear{Eggleton}{1983}]{egg83} Eggleton P. P.  1983, ApJ,
    268, 368
\bibitem[\protect\citeauthoryear{Eggleton \& Kiseleva-Eggleton}{2001}]{egg01} Eggleton P. P., \& Kiseleva-Eggleton L.,  2001, ApJ,
    562, 1012
\bibitem[\protect\citeauthoryear{Gazeas et al.}{2005}]{gaz05} Gazeas K. D., Baran A.,
Niarchos P., Zola S., et al. 2005, Acta Astron., 55, 123
\bibitem[\protect\citeauthoryear{Gurol}{2005}]{gur05} Gurol B., 2005, New Astron., 10, 653
\bibitem[\protect\citeauthoryear{Hut}{1980}]{hut80} Hut P., 1980, A\&A, 92, 167
\bibitem[\protect\citeauthoryear{Hrivnak et al.}{1995}]{hri95} Hrivnak B. J., Guinan E. F.,
\& Lu W. X., 1995, ApJ, 455, 300
\bibitem[\protect\citeauthoryear{Kaluzny}{1990}]{kal90} Kaluzny J.  1990, Acta Astron. 40,
61
\bibitem[\protect\citeauthoryear{Kaluzny \& Shara}{1988}]{kal88} Kaluzny J., \& Shara M. M.,  1988, AJ,
    95, 785
\bibitem[\protect\citeauthoryear{Li et al.}{2004}]{li04} Li L., Han Z., \& Zhang F.,  2004, MNRAS,
    355, 1383
\bibitem[\protect\citeauthoryear{Li et al.}{2005}]{li05} Li L., Han Z., \& Zhang F.,  2005, MNRAS,
    360, 272
\bibitem[\protect\citeauthoryear{Lu \& Rucinski}{1999}]{lu99} Lu W. X., \& Rucinski S. M.,
1999, AJ, 118, 515
\bibitem[\protect\citeauthoryear{Maceroni \& van't Veer}{1996}]{mac96} Maceroni C., \&
van't Veer F., 1996, A\&A, 311, 523
\bibitem[\protect\citeauthoryear{Mateo}{1996}]{mat96} Mateo M.  1996, in {\it The Origins,
Evolutions, and Destinies of Binary Stars in Clusters}, ASP Conf.,
Vol. 90, ed. E. F. Milone \& J. C. Mermilliod, 346
\bibitem[\protect\citeauthoryear{Mateo et al.}{1990}]{mat90} Mateo M., Harris H. C., Nemec J., \& Olszewski E. W.,  1990, AJ,
    100, 469
\bibitem[\protect\citeauthoryear{Ozdemir et al.}{2002}]{ozd02} Ozdemir S., Demircan O.,
Erdem A. et al., 2002, A\&A, 387, 240
\bibitem[\protect\citeauthoryear{Qian et al.}{2005}]{qia05} Qian S., Zhu L., Soonthornthum B., Yuan J., et al., 2005, AJ, 130, 1206
\bibitem[\protect\citeauthoryear{Rasio}{1995}]{ras95} Rasio F. A. 1995, ApJ, 444, L41
\bibitem[\protect\citeauthoryear{Rasio \& Shaprio}{1995}]{ras951} Rasio F. A. \& Shapiro S.
L., 1995, ApJ, 438, 887
\bibitem[\protect\citeauthoryear{Rucinski}{1992}]{ruc92} Rucinski
S. M., 1992, AJ, 104, 1968
\bibitem[\protect\citeauthoryear{Rucinski et al.}{2001}]{ruc01} Rucinski S. M., Lu W. X.,
Mochanacki S. W., Ogloza W., \& Stachowski G., 2001, AJ, 122, 1974
\bibitem[\protect\citeauthoryear{Rucinski \& Vilhu}{1983}]{ruc83}
Rucinski S. W., Vilhu O., 1983, MNRAS, 202, 1221
\bibitem[\protect\citeauthoryear{Vaiana}{1980}]{vai80} Vaiana G.,
1980, in Wayman P.A., ed., Highlights of Astronomy, Vol. 5,
Reidel, Dordrecht, P. 419
\bibitem[\protect\citeauthoryear{Vilhu}{1983}]{vil83} Vilhu O.,
1983, in West R. M., ed., Highlights of Astronomy, Vol. 6. Reidel,
Dordrecht, P. 643
\bibitem[\protect\citeauthoryear{Yakut \& Eggleton}{2005}]{yak05} Yakut K., \& Eggleton P. P., 2005, ApJ,
    629, 1055
\bibitem[\protect\citeauthoryear{Yan \& Mateo}{1994}]{yan94} Yan L., \& Mateo M.,  1994, AJ,
    108, 1810
\bibitem[\protect\citeauthoryear{Yang et al.}{2005a}]{yan05} Yang Y.-G., Qian S.-B., \& Zhu L.-Y., 2005a, AJ, 130, 2252
\bibitem[\protect\citeauthoryear{Yang}{2005b}]{yan051} Yang Y.-G., 2005b, PhD Thesis,
Yunnan Observatory of Chinese Academy Sciences.
\bibitem[\protect\citeauthoryear{Zhang et al.}{2005}]{zha05} Zhang X. B., Zhang R. X., \& Deng
L., 2005, AJ, 129, 979

\end{thebibliography}
\end{document}